\documentclass[sigconf]{acmart}
\usepackage{subcaption}

\graphicspath{ {./images/} }

\copyrightyear{2026}
\acmYear{2026}
\setcopyright{cc}
\setcctype{by-nc-nd}
\acmConference[CHI EA '26]{Extended Abstracts of the 2026 CHI Conference on Human Factors in Computing Systems}{April 13--17, 2026}{Barcelona, Spain}
\acmBooktitle{Extended Abstracts of the 2026 CHI Conference on Human Factors in Computing Systems (CHI EA '26), April 13--17, 2026, Barcelona, Spain}
\acmDOI{10.1145/3772363.3798787}
\acmISBN{979-8-4007-2281-3/2026/04}

\begin{document}

\title{Beyond Privacy Labels: How Users Perceive Different Information Sources for Understanding App's Privacy Practices}

\author{Varun Shiri}
\email{varun.shiri@polymtl.ca}
\orcid{0009-0006-9695-6737}
\affiliation{%
  \department{Department of Computer and Software Engineering}
  \institution{Polytechnique Montreal}
  \city{Montreal}
  \state{QC}
  \country{Canada}
}
\author{Charles Liu}
\email{peiyong.liu@mail.mcgill.ca}
\orcid{0009-0006-8278-2377}
\affiliation{%
  \department{School of Computer Science}
  \institution{McGill University}
  \city{Montreal}
  \state{QC}
  \country{Canada}
  %ORCID: 0009-0006-8278-2377
}
\author{Keyu Yao}
\email{keyu.yao@mail.mcgill.ca}
\orcid{0009-0003-5709-4175}
\affiliation{%
  \department{School of Computer Science}
  \institution{McGill University}
  \city{Montreal}
  \state{QC}
  \country{Canada}
  %ORCID: 0009-0003-5709-4175
}
\author{Jin L.C. Guo}
\email{jin.guo@mcgill.ca}
\orcid{0000-0003-1782-1545}
\affiliation{%
  \department{School of Computer Science}
  \institution{McGill University}
  \city{Montreal}
  \state{QC}
  \country{Canada}
}
\author{Jinghui Cheng}
\email{jinghui.cheng@polymtl.ca}
\orcid{0000-0002-8474-5290}
\affiliation{%
  \department{Department of Computer and Software Engineering}
  \institution{Polytechnique Montreal}
  \city{Montreal}
  \state{QC}
  \country{Canada}
}

\begin{abstract}
Despite having growing awareness and concerns about privacy, technology users are often insufficiently informed of the data practices of various digital products to protect themselves. Privacy policies and privacy labels, as two conventional ways of communicating data practices, are each criticized for important limitations---one being lengthy and filled with legal jargon, and the other oversimplified and inaccurate---causing users significant difficulty in understanding the privacy practices of the products and assessing their impact. To mitigate those issues, we explore ways to enhance privacy labels with the relevant content in complementary sources, including privacy policy, app reviews, and community-curated privacy assessments. Our user study results indicate that perceived usefulness and trust on those information sources are personal and influenced by past experience. Our work highlights the importance of considering various information needs for privacy practice and consolidating different sources for more useful privacy solutions.
\end{abstract}



\maketitle

\section{Introduction}
The emerging data economy has prompted the invasive data practices of many app and service providers~\cite{Zuboff2019}. Data privacy, therefore, now concerns almost every technology user. While users' attitudes and behaviours diversified with the increasingly pervasive and aggravated issue of online privacy~\cite{Dupree2016,Zhang2024}, many users started to have growing concerns about how their data is collected, stored, and used by companies and governments~\cite{EmamiNaeini2023,Song2025}. Some recent efforts also explored design solutions to further motivate users to pay attention to their online privacy~\cite{Shiri_2024}. Despite the growing concern and awareness on this problem, however, users generally felt unable to make informed decisions to protect themselves due to a lack of understanding about the data handling practices performed by the various apps and services they use every day~\cite{knowlesUnParadoxingPrivacyConsidering2023}. This trend is observed in many public opinion surveys conducted worldwide~\cite{EUSurvey2020,CanadaSurvey2025,PewSurvey2023}.
 
Privacy policy, the standard channel of communication between users and service providers, now mandated by many governments, is supposed to inform users of the service providers' privacy practices. In reality, however, the information potentially valuable for users is often hidden in long and hard-to-comprehend legal terms and therefore mostly ignored~\cite{mcdonald2008cost}. As a shortcut to privacy policies, privacy labels were proposed to provide short and uniform descriptors that allow users to quickly learn about how their data is collected and used~\cite{Li_2024}. This practice is now enforced by major mobile app platforms, such as Apple App Store and Google Play Store~\cite{10190677}. However, the reliability and effectiveness of using privacy labels as an instrument to communicate privacy practices are not guaranteed~\cite{kochKeepingPrivacyLabels2022, 10.1145/3563967}. Those seemingly straightforward labels were found to be confusing for both users and developers~\cite{10.1145/1572532.1572538, balash2024would, 10.1145/3491102.3502012}, resulting in the privacy labels being inaccurate and hard to use. Moreover, the information included in privacy labels is limited, usually failing to answer many privacy-related questions that users may have~\cite{zhang2023privacy}.  

In this paper, we aim to support users who are motivated to attend to their online privacy and intend to understand the service providers' data practices, by helping them navigate the jungle of privacy-related information. In particular, we explore a middle ground between the often over-simplified privacy labels and the lengthy privacy policies, through investigating a new design space for tools that augment the privacy labels with sources that can provide users with privacy-related information from different sources representing distinct perspectives. We want to understand how users perceive and adopt those sources when they attempt to comprehend the app's privacy practices. 

To this end, we designed a preliminary prototype with four versions, each of which augments the privacy labels with additional information, including (1) summaries of the privacy policy, (2) excerpts from the privacy policy, (3) privacy-related app reviews, and (4) community-curated assessment from \href{https://tosdr.org/}{ToS;DR} (a community-driven, nonprofit project that summarizes and rates privacy policies and other legal documents of apps and services). We conducted a user study to identify users' feedback regarding each type of information. Our results revealed advantages and risks of each information source. Together, they call for future work that integrates different sources and consolidates efforts from different parties to enhance users' understanding of apps' privacy practices and the corresponding risks.
\section{Prototype Design}

Our prototype is developed as a Chrome plugin for the Android app pages on the Google Play store, although our design elements can be adapted to other app stores. In April 2022, Google introduced mandatory privacy labels in the Play Store~\cite{10190677}. This mandate requires developers to disclose how their apps collect, share, and handle data through privacy labels presented on each app’s Data Safety page.

We designed our prototype to (1) offer additional context and insights to enrich the privacy labels and (2) experiment with various formats and sources to effectively convey the app's privacy practice. Privacy-related information can be derived from different sources. Privacy policies are the main source of this information, but they are often long, unclear, and full of complex jargon~\cite{mcdonald2008cost}. Advancements in automated summarization techniques can help to address these problems~\cite{singh-etal-2024-eros}. Furthermore, community-generated content such as user reviews in app stores and community-driven projects can also shed insights on an application's privacy practices~\cite{Shiri_2024}. Drawing from previous work, we formulated the following four design elements to supplement the information relevant to a specific privacy label:

\begin{figure*}[t]
    \centering
    \begin{subfigure}[b]{0.95\textwidth}
         \centering
         \includegraphics[width=0.75\textwidth]{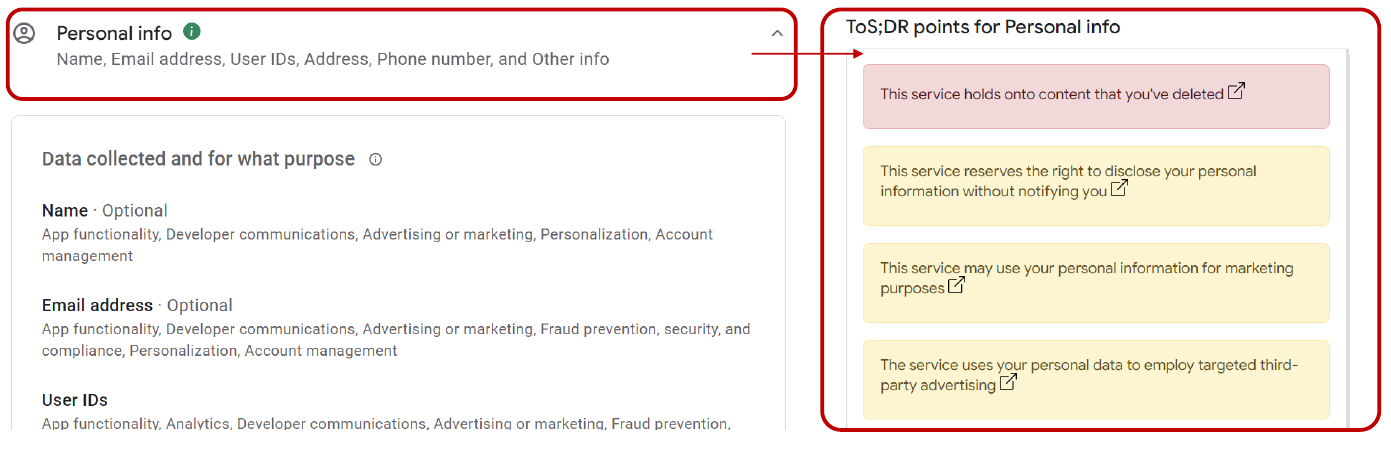}
        \caption{A panel of ToS;DR assessment on how personal information is handled by the app is shown when the user clicks the label on the Data Safety page. Similarly, a panel is shown for each of the other design elements in different prototype versions.}
        \label{figure:preliminary_tosdr}
        \Description{A panel of ToS;DR assessment on how personal information is handled by the app is shown when the user clicks the label of ``Personal Info'' on the Data Safety page. Similarly, a panel is shown for each of the other design elements in different prototype versions.}
     \end{subfigure}
     \vspace{5mm}
     
    \begin{subfigure}[b]{0.3\textwidth}
         \centering
         \includegraphics[width=\textwidth]{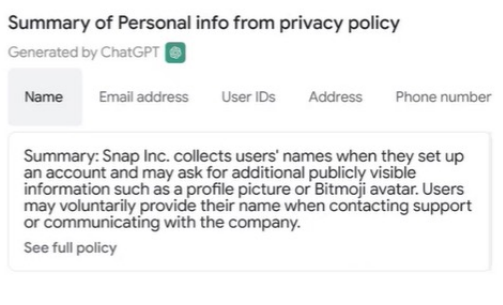}
         \caption{Privacy policy summary panel.}
         \label{figure:preliminary_summary}
         \Description{The screenshot of the design element of the privacy policy summary.}
     \end{subfigure}
     \hfill
     \begin{subfigure}[b]{0.3\textwidth}
         \centering
         \includegraphics[width=\textwidth]{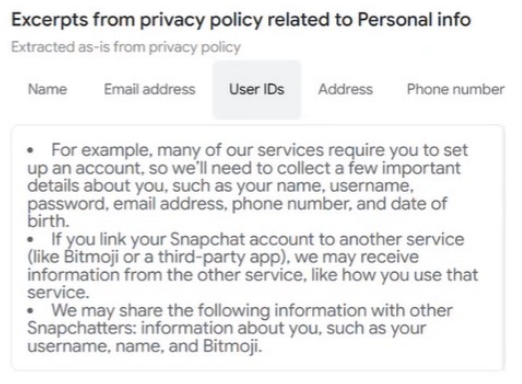}
        \caption{Privacy policy excerpt panel.}
         \label{figure:preliminary_excerpt}
         \Description{The screenshot of the design element of the privacy policy excerpts.}
     \end{subfigure}
     \hfill
     \begin{subfigure}[b]{0.3\textwidth}
         \centering
         \includegraphics[width=\textwidth]{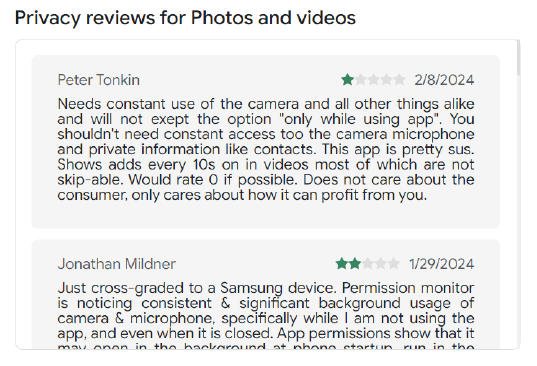}
         \caption{The panel of app review concerning privacy.}
         \label{figure:preliminary_review}
         \Description{The screenshot of the design element of the app review concerning privacy.}
     \end{subfigure}
    \caption{Our preliminary prototype supplements the privacy labels on the Data Safety page.}  
    \label{figure:preliminary_design}
    \Description{Our preliminary prototype supplements the privacy labels on the Data Safety page with one of each design element.}
\end{figure*}

\textit{Privacy Policy Summary.} 
Summaries of the privacy policy provide a quick glance at the relevant portions of the privacy policy for each data type on the Data Safety page (see Fig.~\ref{figure:preliminary_summary}). When the user clicks the privacy label of a certain data category, e.g., personal info, the summary on each data type under that category is located under the tab corresponding to the name of that data type, e.g., name, email address, user IDs, etc. Initial summary was generated using ChatGPT. We then manually verified and revised the generated summaries with the privacy policy. 

\textit{Privacy Policy Excerpts.}
Excerpts are sentences extracted verbatim from the privacy policy that are considered relevant to particular data types on the Data Safety page (see Fig.~\ref{figure:preliminary_excerpt}). In the prototype, we manually selected excerpts relevant to each data type by reading the privacy policy of the target app and displayed them as a bullet list. Excerpts often cover more than one data type. In such cases, they were put under the tab of each mentioned data type.

\textit{App Reviews Concerning Privacy.}
App reviews provide valuable insights into users' attitudes and experiences, serving as an indicator of app quality~\cite{10.1145/2414536.2414577, 10.1145/2950290.2950299}, as well as the privacy practice of the app~\cite{Shiri_2024}. Reviews relevant to the corresponding privacy labels were extracted and displayed in reverse chronological order (see Fig.~\ref{figure:preliminary_review}). 

\textit{ToS;DR Assessment.}
\href{https://tosdr.org/}{ToS;DR} is a community-driven project that provides summarized versions of terms of service and privacy policies for various platforms and assigns a grade to each application denoting its privacy practice. The different clauses within the privacy policy or terms of service are classified by community members into cases and assigned a colour code: a grey colour indicates a neutral privacy effect, green a beneficial practice, red a potentially harmful one, and yellow a mildly concerning practice. We selected the ones that are relevant to the corresponding data categories and displayed them based on the risk level, with the most risky ones on top (see Fig.~\ref{figure:preliminary_tosdr}).

In our prototype, we created four different versions, each containing one design element. When users click on a privacy label on the Data Safety page, they can access one of the design elements, as shown in Fig.~\ref{figure:preliminary_tosdr}. In this study, as we aim to understand how these sources of information can support user comprehension of app's privacy practices, we performed manual extraction of information. Automated techniques can be explored in future work.

\section{User Study Methods}
We conducted a user study with the prototype to explore users' feedback on each design element. The study was approved by the research ethics boards of all involved universities. Participants were recruited through social network and personal connections, including three females and seven males. Their ages ranged from 20 to 60 years, with a median age group of 30-40. Participants had diverse occupations, including students and professionals in software development, medicine, law, management, and professional writing.

Each study lasted around one hour and the participant was compensated with 30 CAD. During the study, participants were first asked about their general experiences and attitudes toward privacy policies. We then presented the participants with the Data Safety page augmented with each of the four design elements. For the user study, we selected \textit{Snapchat}, a popular messaging app that allows users to exchange pictures and videos, as an example app. We chose this app due to its popularity, rich privacy-related content, and identified discrepancies between its Data Safety page and privacy policy. Participants were given time to explore and interact with the four design elements of the prototype; the order in which participants engaged with these elements was counterbalanced using Latin Square to minimize order effects. Participants were then asked targeted questions to assess their experiences, likes and dislikes. Questions focused on ease of use, clarity, trustworthiness, and overall satisfaction with the feature. The sessions were recorded and fully transcribed. We conducted a thematic analysis to analyze the data.

\section{User Study Results}
\label{subsec:user_study_1_result}
Our study revealed an extensive set of user perceptions regarding privacy information from different sources. 

\paragraph{\textbf{Summary makes complex information more understandable but poses trust issues.}}
Participants appreciated the clarity and simplicity of the privacy summaries. For example, P3 stated that \textit{``the descriptions in full sentences is way easier for the layman to understand.''} The conciseness of the summaries was also appreciated, as P1 said, \textit{``3 to 4 lines, it's good''}. Moreover, some participants thought that the summaries had a good level of elaboration to complement the privacy labels. P3 highlighted the value of the additional information, \textit{``it does provide a lot more information in the summary ... I wasn't able to understand anything from the previous page [the Data Safety page].} Additionally, the summaries based on the privacy labels were recognized for their well-organized categorization, making it easier for users to find relevant information. P5 shared that you \textit{``[get] a little more information about each individual setting''}.

While acknowledging the usefulness of such summaries as an initial point of reference, most participants stated that they did not trust AI-generated summary. Some participants suspect AI-generated content might miss useful information. As P10 put it: \textit{``Some technical information I found was maybe missing to get beyond what I could sort of guess from this very, very concise summary.''} Others shared concerns about the accuracy of the information, as P1 stated, \textit{``Even though you know that the models are good, they are not as reliable as a human because it doesn’t understand the intricacies.''} Most participants' concerns about AI's accuracy stemmed from their past personal experiences.

\paragraph{\textbf{Excerpts from privacy policy provide granular and authoritative content but might cause confusion and information overload.}} 
Participants found the excepts from the privacy policy organized by privacy labels useful for scanning information quickly. For example, P4 felt that the granularity of the excerpts, present for each of the data types, contributed to a \textit{``deeper understanding of the [privacy policy] summary tab.''} 
Some participants found the excerpts more trustworthy than the summary as they were directly extracted from the policy rather than being generated. 
For example, P10 pointed out:  {``I really value the authoritativeness of information and the sort of trustworthiness of the information.''} P10 further suggested that excerpts provided more evidence and examples than summarized contents, {``... the excerpt that definitely gives me more information and I attempt to trust this information a bit more.''}

At the same time, some participants found that the excerpts required additional reading, which in some cases might lead to ``\textit{information exhaustion}'' (P9). P1 also pointed out, \textit{``the only thing that worries a user would be the amount of information you are exposing.''} Some participants felt that the excerpts also lacked some contextual information either due to how excerpts were selected or the incompleteness of the privacy policy itself. As P2 mentioned, \textit{``I would like if it had more information, like ... if I uninstall the app for instance, does it revoke these permissions automatically, or does it not? This is stuff that I would love to see, because that would give me way more confidence to download the app.''}

\paragraph{\textbf{App reviews offer relatable real-user experiences but might not be reliable.}} Several participants appreciated the filtered privacy-specific reviews that were categorized by the data types. Real user experiences and the specific issues described in the reviews helped the participants relate to the content, as P3 shared that they appreciated the reviews from \textit{``people who are using the app just like me. I think it's a lot easier to understand the things that they're saying because they write in a more absorbable format.''} P2 stated about the reviews: \textit{``this version of the data explanation gives a more concrete example of what that policy translates to in terms of a user experience...''} Other participants also mentioned that the chronological order of the reviews, sorted by the latest reviews, was helpful. 

The validity of the review was not agreed on by the participants. Some found the reviews trustworthy as those reviews were the experiences of individuals who tried and tested the app, as echoed in P8's words: \textit{``These are from real people, real story. So I have no reason not to trust it.''}. At the same time, many participants remarked that the reviews might be skewed toward the negative, as P2 commented: \textit{``I've never seen anyone writing a glowing 5 star review about the privacy policy of an app.''} Some participants highlighted different aspects about the reviews that might undermine their trust in this source of information. P10 expressed that \textit{``when a review gets super extreme, I trust it less because it seems kind of emotional in the sense that it's a less rational discussion of the pros and cons.''} Participants also questioned the validity of the reviews, because \textit{``there's no vouching process for these reviewers''} (P3). Finally, P6 voiced concerns about applications paying to get favourable reviews and reviewers lacking sufficient knowledge.

\paragraph{\textbf{ToS;DR provides
straightforward risk indicators with crowd-sourced insights but is perceived as unreliable.}} 

The colour coding of ToS;DR indicating the level of privacy risk was greatly appreciated by some participants. For example, P3 consider the colour coding a clear indication of their privacy risk \textit{``if I was more concerned about financial info, if I see a lot of red here instantly, I would already rule out this app.''} P2 especially liked the privacy grade bar at the top of the page, recognizing its value to get a quick overview of the privacy practice of an app: \textit{``If anything, it alerts you to the fact that there may be something you don't like here, and it might be enough motivation for you to go and investigate further.''} Participants also found the crowd-sourced information to be more trustworthy than the reviews, in general, as the content in ToS;DR is moderated and \textit{``gives an element of social proof, it makes it feel a little more legitimate.''} (P2). Some participants found it beneficial to follow the links leading to the ToS;DR website, which contained further discussion on the assessment. P5 expressed: \textit{''The fact that you can click on this and it takes you to a website that kind of breaks it down and people talking about it, I think that's a good thing''} Others liked the level of detail in the information, as P3 felt that \textit{``these points are very summarized, but also like give a lot of information.''}. 

Some participants perceived ToS;DR information as unreliable sometimes, similar to app reviews and other crowd-sourced information. As P6 highlighted: \textit{``Part of why I trust it is that it's crowd-sourced, but I recognize that crowd-sourcing can be problematic ... In the same way that, like Yelp or TripAdvisor, any review sites are sort of scammy a bit.''} P1 also shows these concerns on the information being biased: \textit{``But I don't think it's completely reliable because there might be a chance that some users might be forced to push that information. They are more inclined towards some of the policies or some of the apps and then they are putting what they believe in it, [and therefore] can be biased as well.''} P2 echoed this point by suggesting that, \textit{``if you find an app that doesn't have X million number of downloads, there's gonna be way less community source information.''}

\section{Discussion and Conclusion}
Users' feedback on each design element illustrates a complex picture of their preferences. The perceived usefulness and trust in each type of information are highly personal and are influenced by their past experience with that type of information. These findings revealed important considerations concerning not only the nature of different information, but also how to integrate them to make the tool more usable and effective.

\textbf{Careful integration of information is needed to prevent information overload and redundancy.} 
Different types of information provide different levels of detail; therefore, its integration needs to be carefully designed so that the information relevant to the user's need at a certain time is accessible. For example, in our prototype, we added the ``See full policy'' button under the summary. This feature was well-received because it allowed the participants to prompt the users to manually verify the summary or find additional information when needed. 
On the other hand, some design elements were considered sub-optimal. The panel in our prototype was designed according to how the Data Safety page was organized, i.e., first by data category, then by data type. For the policy summary and excerpt, the data types were used as separate tabs in the panel. Since the privacy policy often describes similar data types together in the same sentence, we duplicated the sentence under each data type tab (see Fig.~\ref{figure:preliminary_excerpt} and Fig.~\ref{figure:preliminary_summary}). This is complained by some participants, as P7 noted: \textit{``I click sometimes to see the same sentence five times. It's redundant.''}

\textbf{Users' information acquisition strategy balances perceived reliability and familiarity of the sources.} 
Among the four information sources, ToS;DR is the least familiar to most participants. Despite of providing validated assessments to identify issues and loopholes in privacy policies and straightforward colour coding of privacy-associated risks, participants underestimated ToS;DR's reliability due to the lack of familiarity. Being the most familiar information source to users, app reviews were also frequently dismissed by the participants, as participants interpreted product reviews as biased in general. Interestingly, although most participants have never read and thus are unfamiliar with privacy policies, they considered privacy policies as the most authoritative. Participants' perception of different sources suggests an intricate balance between perceived reliability and familiarity when choosing the source to obtain privacy-related information. Future tools should consider features to indicate the reliability and trustworthiness of information sources to help users make informed decisions.

\textbf{Raising users' privacy awareness and knowledge requires consolidating endeavour from multiple parties.}
In the end, the responsibility of providing useful and usable privacy-related information is distributed among different parties, each with its own advantages and risks. The app and service providers are responsible for providing the most accurate information about how they handle user data. However, they may face conflicts of interest in clearly explaining this information and their privacy policies are often hard to understand~\cite{kochKeepingPrivacyLabels2022}. App hosting platforms, such as the Google Play Store and Apple App Store, have the intention of allowing users to quickly digest the apps' privacy practices and sustaining the ecosystem of the mobile app market. However, the privacy labels they reinforced have become a chore for developers and sometimes deviate from the apps' actual practices~\cite{10.1145/3491102.3502012}. Similarly, the user communities often have first-hand information about what the users are concerned about, but their comments and experiences shared in user reviews and forums are frequently deemed unreliable. Thus, tools integrating different sources of information can serve as mediators of these different parties, balancing the advantages and drawbacks of each source and synthesizing the multifaceted information to raise users' privacy awareness and knowledge.

In conclusion, this work investigated how the privacy labels can be supplemented using different types of sources, including the privacy policy, app reviews, and risk assessment from the community. Results from a user study indicated the complementary nature of those sources and the consideration when integrating them during users' comprehension of mobile app privacy practices. Our study points to the importance of considering privacy as a distributed effort and calls for future work on designing tools that consolidate those efforts and ultimately encourage users' informed decisions concerning privacy.
\balance

\bibliography{reference}
\bibliographystyle{ACM-Reference-Format}

\end{document}